\documentclass[aps,a4paper,11pt, nofootinbib]{revtex4}

\usepackage{bm}
\usepackage{mathrsfs}
\usepackage{xcolor,color,graphicx,graphics}
\usepackage[all]{xy}
\usepackage{epsfig,subfigure}
\usepackage{latexsym,amssymb,amsmath,amsfonts} 
\usepackage[english]{babel} 
\usepackage[OT1]{fontenc}
\usepackage[latin1]{inputenc}
\usepackage{makeidx}
\usepackage{hyperref}
\usepackage{color,graphicx,graphics,wrapfig,epsf}

\definecolor{red}{rgb}{1,0,0}

\def\+{^\dagger}

\def\<{\leftarrow}
\def\>{\rightarrow}

\def\({\left(}
\def\){\right)}

\newcommand{\bi}{\begin{itemize}} 				\newcommand{\ei}{\end{itemize}}
\newcommand{\benu}{\begin{enumerate}} 		\newcommand{\enu}{\end{enumerate}}
\newcommand{\bd}{\begin{dinglist}{0}}     \newcommand{\ed}{\end{dinglist}}
\newcommand{\bfig}{\begin{figure}[htbp]}  \newcommand{\efig}{\end{figure}}
        			
\newcommand{\bc}{\begin{center}} 				  \newcommand{\ec}{\end{center}}
\newcommand{\be}{\begin{equation}} 				\newcommand{\ee}{\end{equation}}
\newcommand{\bsub}{\begin{subequations}}  \newcommand{\esub}{\end{subequations}}
\newcommand{\ben}{\begin{eqnarray}} 			\newcommand{\een}{\end{eqnarray}}
\newcommand{\ba}[1]{\begin{array}{#1}} 		\newcommand{\ea}{\end{array}}
\newcommand{\bea}{\begin{equation}\begin{array}{rcl}}
\newcommand{\eea}{\end{array}\end{equation}}

\begin{document}
\title{Ricci-Based Gravity theories and their impact on Maxwell and nonlinear electromagnetic models}

\author{Adria Delhom }
\email{adria.delhom@uv.es}
\affiliation{Departamento de F\'{i}sica Te\'{o}rica and IFIC, Centro Mixto Universidad de
Valencia - CSIC. Universidad de Valencia, Burjassot-46100, Valencia, Spain}
\author{Gonzalo J. Olmo} \email{gonzalo.olmo@uv.es}
\affiliation{Departamento de F\'{i}sica Te\'{o}rica and IFIC, Centro Mixto Universidad de Valencia - CSIC.
Universidad de Valencia, Burjassot-46100, Valencia, Spain}
\affiliation{Departamento de F\'isica, Universidade Federal da
Para\'\i ba, 58051-900 Jo\~ao Pessoa, Para\'\i ba, Brazil}
\author{Emanuele Orazi} \email{orazi.emanuele@gmail.com}
\affiliation{ International Institute of Physics, Federal University of Rio Grande do Norte,
Campus Universit\'ario-Lagoa Nova, Natal-RN 59078-970, Brazil}
\affiliation{Escola de Ciencia e Tecnologia, Universidade Federal do Rio Grande do Norte, Caixa Postal 1524, Natal-RN 59078-970, Brazil}

\date{\today}
\begin{abstract}
We extend the correspondence between metric-affine Ricci-Based Gravity theories  and General Relativity (GR) to the case in which the matter sector is represented by linear and nonlinear electromagnetic fields. This complements previous studies focused on fluids and scalar fields. We establish the general algorithm that relates the matter fields in the GR and RBG frames and consider some applications. In particular, we find that the so-called Eddington-inspired Born-Infeld gravity theory coupled to Maxwell electromagnetism is in direct correspondence with GR coupled to Born-Infeld electromagnetism. We comment on the potential phenomenological implications of this relation. 

\end{abstract}

\maketitle

\section{Introduction}

Among the various families of modified theories of gravity currently available in the literature, the so-called Ricci-Based Gravity theories (RBGs for short) have very peculiar properties that make them particularly interesting. These theories are formulated assuming that metric and connection are independent geometric fields, taking the gravity Lagrangian to be a function of the (inverse) metric and the Ricci tensor of the connection, and coupling the matter sector minimally to the metric $g_{\mu\nu}$. From the field equations, one then finds a direct algebraic relation between the metric $g_{\mu\nu}$ and the stress-energy tensor of the matter fields and, as a result, the metric becomes sensitive to both the total amounts of energy (integration over the sources) and also to the local energy-momentum distributions. This local dependence on the local energy-momentum densities manifests itself through nonlinearities in the matter sector, which generate self-interactions and new couplings among all elementary particles {that can be seen as effective interactions below a high-energy scale $\Lambda_Q$. These effects have been recently used to place the strongest constraints available on the RBG class of models \cite{Latorre:2017uve,PenuelasMirallesDelhom}.}

The emergence of nonlinearities in the matter sector can also be relevant in astrophysical scenarios \cite{Olmo:2019qsj}. In particular, the extreme conditions that exist in the interior and neighborhood of neutron stars has been used to study potential deviations from Maxwell's electrodynamics in the generation and propagation of electromagnetic radiation coming from such sources. In fact, over astrophysical and cosmological distances, nonlinear effects could add up and lead to observable effects on the propagation speed and polarisation of electromagnetic waves. Moreover, with the advent of gravitational wave astronomy and the possibility of multimessenger analyses of neutron stars collisions, the exploration of subtle effects in electromagnetic fields will become closer to observational and experimental reach in the coming years \cite{Abbott:2017oio,TheLIGOScientific:2017qsa,Stratta:2017igm,Williams:2017ibl,Howell:2018nhu,Barack:2018yly,Dai:2019pgx,Barbieri:2019sjc,Tarrant:2019gqg}, complementing in this way ongoing efforts in accelerator experiments. It is thus of utmost importance to scrutinize from a theoretical perspective the influence that modified gravitational dynamics could have on the generation and propagation of electromagnetic waves in strong gravity scenarios. Our purpose here is to elaborate in this direction.

 In this work we continue a program initiated in \cite{Afonso:2018bpv} in which the field equations of Ricci-Based Gravity theories (RBGs) coupled to standard matter are put into correspondence with the field equations of GR coupled to the same matter species but with a different, typically nonlinear Lagrangian. In \cite{Afonso:2018hyj} the focus was on the case of scalar fields and the non-perturbative results obtained there were used to generate new solutions in nonlinear RBG theories starting with known solutions in GR \cite{Afonso:2019fzv}. In \cite{Afonso:2018mxn} the equivalence between (linear or nonlinear) electric fields  and anisotropic fluids was exploited in a similar fashion. In particular, it was shown there that the so-called Eddington-inspired Born-Infeld gravity model (EiBI) coupled to Maxwell electrodynamics is in correspondence with GR coupled to Born-Infeld nonlinear electrodynamics. Whether this result also extends to all kinds of electromagnetic fields is not known, since the approach of  \cite{Afonso:2018mxn} is limited to electric fields, i.e., without magnetic component. To clarify this point and set the path to explore more general electromagnetic effects in nonlinear RBGs, it is necessary to extend the analysis of \cite{Afonso:2018mxn} to the complete electromagnetic case. Paralleling the approach of \cite{Afonso:2018hyj} for scalar fields, here we extend the correspondence between GR and RBGs for arbitrary (linear and nonlinear) electromagnetic fields. Our results confirm that a correspondence for electromagnetic fields also exists, and recovers the electric case in the appropriate limit. Though our results are valid for any RBG coupled to (linear or nonlinear) electromagnetic fields, we consider two particular examples for simplicity, namely, a quadratic $f(R)$ model and the EiBI gravity theory coupled to Maxwell electrodynamics. Interestingly, in the particular case of EiBI coupled to Maxwell, in the GR frame one exactly obtains Born-Infeld electrodynamics, which puts forward an unexpected deep relation between both theories. Let us mention that {a perturbative version of this technique was already employed for spin 1/2 fields in \cite{Latorre:2017uve} to observationally constrain RBG models. The difficulty in developing the full non-perturbative correspondence for spin 1/2 fields stems from the fact that spinor fields source a torsion term in the affine connection, which introduces technical difficulties that are not present in the bosonic case. Although neglecting torsion was physically justified for the perturbative approach in \cite{Latorre:2017uve}, there is ongoing work in formulating a full non-perturbative correspondence for fermionic matter.}

 The paper is organized as follows. In section II we explain how to write the field equations associated to the metric variation in terms of the Einstein tensor depending on an auxiliary metric. In order to recover the Einstein field equations, specific conditions are presented that relate the matter sectors of the GR and RBG representations yielding the fundamental equations that describe the relations between them. In subsection IIA the deformation matrix that represents an efficient tool to express the auxiliary metric in terms of the original one is introduced. In section III it is shown that the equations describing the correspondence and the deformation matrix can be recast in terms of the fundamental invariants of the electromagnetic field. In sections IV and V, explicit examples that implement the correspondence are presented for gravity sectors of $f(R)$ and Eddington-inspired Born-Infeld gravity theory.

\section{Field equations of RBGs} \label{sec:I}

The gravity theories we will be dealing with in this work, RBGs, can be defined as the most general class of theories which enjoy diffeomorphism and projective symmetries and the Lagrangian is a function of the metric and the Ricci tensor of an independent connection. Thus the RBG class of theories is described by the general action
\be\label{action}
S=\int d^4x\sqrt{|g|}\left[{\cal L}_G\left(g_{\mu\nu},R_{(\mu\nu)}(\Gamma)\right)+{\cal L}_m\left(g_{\mu\nu},\psi\right)\right] \ ,
\ee

where $R_{(\mu\nu)}(\Gamma)$ is the symmetrized\footnote{Under a projective transformation, the symmetrized Ricci tensor remains invariant, while its antisymmetric part transforms non-tirvially. Thus the requirement of projective symmetry allows only the symmetrized Ricci tensor to appear in the action, which ensures the stability of the theories \cite{BeltranJimenez:2019acz}.} Ricci tensor, $R_{\mu\nu}(\Gamma)=R^\alpha{}_{\mu\alpha\nu}(\Gamma)$, and ${\cal L}_m$ represents the matter Lagrangian, with $\psi$ denoting collectively any bosonic matter fields.  Since we are assuming that metric and connection are independent objects, the field equations will follow by independent variations with respect to $g^{\mu\nu}$ and $\Gamma^\alpha{}_{\mu\nu}$. The field equation associated to the variation of the connection of these theories were carefully studied in \cite{Afonso:2017bxr} to explore in detail the role of torsion, and a general analisys of the matthematical formalism of RBGs can be found in \cite{BeltranJimenez:2017doy}. Here we follow a different but complementary approach, mainly focussing on the metric variation and showing that the introduction of the auxiliary metric used to solve the connection field equation in \cite{Afonso:2017bxr} is also natural when trying to recover Einsteins equations from the metric equations of RBGs. To show this, let us consider the metric variation of the RBG action (\ref{action}), which leads to the following metric field equations:
\be
g^{\mu\rho}\frac{\delta{\cal L}_G\left[g,R(\Gamma)\right]}{\delta g^{\rho\nu}}-\frac12 {\cal L}_G\left[g,R(\Gamma)\right]\delta^\mu_\nu = \frac12 T^\mu{}_\nu \ ,\label{MetricfieldEq}
\ee
where, the matter stress-energy tensor is defined as usual
\be
T_{\mu\nu}\equiv -\frac{2}{\sqrt{|g|}}\frac{\delta\left[\sqrt{|g|}{\cal L}_m(g,\psi)\right]}{\delta g^{\mu\nu}}\ \ . \label{EnergyMomentum}
\ee
Given that  ${\cal L}_G$ must be a scalar under diffeomorphism and projective transformations, the gravity Lagrangian must be a function of powers of traces of the tensor $g^{\mu\rho}R_{(\rho\nu)}$, i.e. we have that the dependence of the gravity Lagrangian on the metric and the Ricci tensor must be of the form ${\cal L}_G[g^{\mu\rho}R_{(\rho\nu)}]$. This dependence leads to the following relation
 \be
\frac{\delta{\cal L}_G\left[g^{\mu\delta}R_{\delta\gamma}\right]}{\delta g^{\rho\nu}}=\frac{\delta{\cal L}_G\left[g^{\mu\delta}R_{\delta\gamma}\right]}{\delta R_{\alpha\sigma}}g_{\alpha\rho}R_{\sigma\nu}\,,
\ee
which allows us to trade the variation of the gravity Lagrangian with respect to the metric in(\ref{MetricfieldEq}) for the variation with respect to the Ricci tensor. Using this property, the equations of motion assume the following form
\be
\frac{\partial{\cal L}_G}{\partial R_{\mu\rho}} R_{\rho\nu}(\Gamma) = \frac12 T^\mu{}_\nu + \frac12 {\cal L}_G\delta^\mu_\nu\,.\label{PreEinstein}
\ee
Having in mind the form of the Einstein tensor, it is natural to define an auxiliary metric $q_{\mu\nu}$ as
\be
\sqrt{|q|}q^{\mu\nu}=2\kappa^2\sqrt{|g|}\frac{\delta{\cal L}_G}{\delta R_{\mu\nu}}\ \ ,
\label{q}
\ee
where $q^{\mu\alpha} q_{\alpha\nu}=\delta^\mu{}_\nu$ defines $q_{\mu\nu}$ from the above equation\footnote{{For consistency, we must require that the derivative of the Lagrangian with respect to the Ricci tensor is invertible, as is always the case for UV modifications of GR within the RBG class of theories}}. In terms of the new metric,  \eqref{PreEinstein} can be written as
\be
q^{\mu\rho}R_{\nu\rho}(\Gamma) = \kappa^2 \sqrt{\frac{|g|}{|q|}}\left(T^\mu{}_\nu + {\cal L}_G\delta^\mu_\nu\right) \ ,
\ee
which can be manipulated algebraically to obtain
\be
q^{\mu\rho}R_{\nu\rho}(\Gamma) - \frac12 q^{\rho\sigma}R_{\rho\sigma}(\Gamma) \delta^\mu_\nu= \kappa^2 \sqrt{\frac{|g|}{|q|}}\left\{T^\mu{}_\nu - \left[{\cal L}_G +\frac12 T\right]\delta^\mu_\nu\right\}\,.\label{EE}
\ee
Notice that when the gravity Lagrangian is ${\cal L}_G=\frac{1}{2\kappa^2}R$, this equation properly recovers the Einstein-Palatini version of GR with the auxiliary metric as the corresponding metric.  For notational convenience, we will denote the left-hand side of (\ref{EE})  as ${\mathcal{G}^\mu}_\nu(q,\Gamma)$. Substituting the explicit expression of the energy momentum tensor in terms of the matter Lagrangian
\be
{T^\mu}_\nu=g^{\mu\alpha}T_{\alpha\nu}={\cal L}_m(g_{\mu\nu},\psi)\delta^\mu{}_\nu-2g^{\mu\rho}\frac{\delta{\cal L}_m(g_{\mu\nu},\psi)}{\delta g^{\rho\nu}}\,,\label{T0}
\ee
 \eqref{EE}  become
\be
{\mathcal{G}^\mu}_\nu(q,\Gamma)= -\kappa^2 \sqrt{\frac{|g|}{|q|}}\left[2g^{\mu\rho}\frac{\delta{\cal L}_m(g_{\mu\nu},\psi)}{\delta g^{\rho\nu}}+\delta^\mu_\nu\left(\mathcal{L}_G+\mathcal{L}_m(g_{\mu\nu},\psi)-g^{\rho\sigma}\frac{\delta{\cal L}_m(g_{\mu\nu},\psi)}{\delta g^{\rho\sigma}}\right)\right]
\,.\label{EE1}
\ee
Usually,  \eqref{EE1} is obtained by using the field equations for the connection given by \eqref{action}, and thus on-shell with respect to the connection. However, it is worth stressing that here \eqref{EE1}  is off-shell with respect to the connection equation. Note that, since $\mathcal{L}_G$ can be written as a function of the stress-energy tensor on-shell, \eqref{EE1} is formally equivalent to the metric equations for GR in the Palatini approach with a suitably modified stress-energy tensor. 

\subsection{Mapping RBGs into GR}\label{sec:MappingFromTmn}

The structure of the field equations \eqref{EE1} is formally identical to that found in Einstein's theory, though the left-hand side is referred to the metric $q_{\mu\nu}$ while the right-hand side contains the metric $g_{\mu\nu}$. Thus, though this representation might be notationally convenient, from a practical perspective it entails important complications due to the need to relate both types of metrics in order to find explicit solutions. It is thus desirable to see if, by  a suitable field redefinition, the right-hand side could be written as the stress-energy tensor of a matter theory coupled to the metric $q_{\mu\nu}$. If this is possible, then the field equations of our modified gravity theory would become equivalent to a problem in the framework of GR. Note, in this sense, that the relation between $q_{\mu\nu}$, $g_{\mu\nu}$, and the matter fields is algebraic. In other words, we want to determine if (\ref{EE1}) can be written as
\be
{\mathcal{G}^\mu}_\nu(q,\Gamma) = \kappa^2 \left( -2{q}^{\mu\rho}\frac{\delta\tilde{\cal L}_m({q}_{\mu\nu},\psi)}{\delta {q}^{\rho\nu}}+\delta^\mu_\nu\tilde{\mathcal{L}}_m({q}_{\mu\nu},\psi)\right)\, , \label{GR EE}
\ee
where $\tilde{\mathcal{L}}_m({q}_{\mu\nu},\psi)$ represents a modified matter Lagrangian {minimally} coupled to the auxiliary metric ${q}_{\mu\nu}$ which defines the matter Lagrangian of Einstein frame representation of the corresponding RBG. If this correspondence exists, then the relation between the matter sectors in the two frames has to be a solution of the following equations
\be
\begin{split}
\delta^\mu_\nu\left(g^{\rho\sigma}\frac{\delta{\cal L}_m(g_{\mu\nu},\psi)}{\delta g^{\rho\sigma}}-\mathcal{L}_G-\mathcal{L}_m(g_{\mu\nu},\psi)\right)&-2g^{\mu\rho}\frac{\delta{\cal L}_m(g_{\mu\nu},\psi)}{\delta g^{\rho\nu}}
=\\
&\sqrt{\frac{|q|}{|g|}} \left( \delta^\mu_\nu\tilde{\mathcal{L}}_m({q}_{\mu\nu},\psi)-2{q}^{\mu\rho}\frac{\delta\tilde{\cal L}_m({q}_{\mu\nu},\psi)}{\delta {q}^{\rho\nu}}\right)\ .\label{mapping}
\end{split}
\ee
This equation will be fundamental for the subsequent discussions in this paper, as it will allow us to construct the matter Lagrangian of the GR representation in terms of the original matter Lagrangian in the RBG representation once a particular gravity Lagrangian of the RBG class is specified. Remarkably, the derivation presented here shows that the correspondence between {the metric equations of RBGs and GR established by (\ref{mapping}) is independent of the connection equation, which represents a novelty with respect to previous works \cite{Afonso:2018bpv}. This implies a full dynamical correspondence between RBGs and GR when formulated in the metric-affine approach in the sense that the field equations for both metric and connection are formally equivalent. As it is well known, metric-affine GR yields standard GR (with possible torsion terms if includding fermions) after solving the connection equation and plugging the solution back into the metric field equations. }

\subsection{The deformation matrix}

In order to solve the mapping equations introduced above, it is convenient to the deformation matrix $\hat\Omega$, which relates $g_{\mu\nu}$ and $q_{\mu\nu}$  as 
\be
q_{\mu\nu}=g_{\mu\rho}\Omega^\rho{}_\nu\,.\label{Om}
\ee
This new object together with the above relation allows to rephrase the definition of the auxiliary metric equation \eqref{q}  as
\be
\sqrt{\Omega}\left(\Omega^{-1}\right)^\mu{}_\nu = 2\kappa^2 \frac{\partial{\cal L}_G}{\partial R_{\mu\rho}} g_{\rho\nu}\,.\label{OmG}
\ee
where $q^{\mu\nu}=\left(\Omega^{-1}\right)^\mu{}_\rho g^{\rho\nu}$ has been used. Taking the determinant of both sides of \eqref{OmG}, one finds
\be
\Omega = \left(2\kappa^2\right)^4\det{\left(\frac{\partial{\cal L}_G}{\partial R_{\mu\rho}} g_{\rho\nu}\right)}\,.\label{DetOm}
\ee
so that the inverse deformation matrix reads
\be
\left(\Omega^{-1}\right)^\mu{}_\nu = \frac{1}{2\kappa^2\sqrt{\det{\left(\frac{\delta{\cal L}_G}{\delta R_{\alpha\lambda}}g^{\lambda\beta}\right)}}}\frac{\partial{\cal L}_G}{\partial R_{\mu\rho}} g^{\rho\nu}\,.\label{InvOm}
\ee
At this point it is important to note that according to (\ref{PreEinstein}), the Ricci tensor $R_{\mu\nu}(\Gamma)$ and all its possible scalar contractions with $g^{\alpha\beta}$ can be algebraically related to ${T^\mu}_\nu$. As a result, the matrix $\hat\Omega$ turns out to be a function of ${T^\mu}_\nu$, which implies that the relation between $q_{\mu\nu}$ and $g_{\mu\nu}$ defined in (\ref{Om}) will be highly nonlinear in general. However, notice that $\hat\Omega$ can always be expanded as a power series of the stress-energy tensor as
\be
\Omega^\mu{}_\nu=\sum_{n=0}^\infty \frac{a_n}{\Lambda_Q^{4n}}(T^n)^\mu{}_\nu\,,\label{Om-T-Exp}
\ee
where $(T^n)^\mu{}_\nu\equiv T^\mu{}_{\lambda_2}T^{\lambda_2}{}_{\lambda_3}\cdots T^{\lambda_{n}}{}_{\nu}$, and where $\Lambda_Q$ is a high-energy scale that characterizes the new interaction terms induced in the matter sector in the Einstein frame. This relation will be useful later in our explicit construction of the Lagrangian $\tilde{\mathcal{L}}_m({q}_{\mu\nu},\psi)$ which maps our original RBG theory with a matter Lagrangian ${\mathcal{L}}_m({g}_{\mu\nu},\psi)$ minimally coupled to $g_{\mu\nu}$ into GR with a matter Lagrangian $\tilde{\mathcal{L}}_m({q}_{\mu\nu},\psi)$ {minimally} coupled to $q_{\mu\nu}$ and the same matter content with new interaction terms, though new symmetries may arise\footnote{{Notice that since the new interactions will arise through the dependence of $\Omega^\mu{}_\nu$ on the mater stress-energy tensor, they will feature at least the same symmetries as the original matter action, since the stress-energy tensor does.}.}.

\section{RBG and Einstein frame variables for NEDs}\label{sec:RBG_NEDs}

In the following, we will assume that the matter sector is represented by a general nonlinear electrodynamics theory (NED) described by a Lagrangian  ${\mathcal{L}}_m({g}_{\mu\nu},A_\mu)$ that depends on the metric only through the electromagnetic invariants $K = -\frac12 F_{\mu\nu}F^{\mu\nu}$ and  $G=\frac14 F_{\mu\nu}{}^\star F^{\mu\nu}$,  with usual definitions for the photon field-strength $F_{\mu\nu}=2\partial_{[\mu}A_{\nu]}$ and its Hodge dual ${}^\star F^{\mu\nu}=\frac12 \epsilon^{\mu\nu\alpha\beta}F_{\alpha\beta}$. For such Lagrangians, one finds
\be
g^{\mu\rho}\frac{\delta{\cal L}_m}{\delta g^{\rho\nu}}=g^{\mu\rho}\frac{\partial{\cal L}_m}{\partial K}\frac{\delta K}{\delta g^{\rho\nu}} + g^{\mu\rho}\frac{\partial{\cal L}_m}{\partial G}\frac{\delta G}{\delta g^{\rho\nu}} = K^\mu{}_\nu\frac{\partial{\cal L}_m}{\partial K}+\frac12\delta^\mu_\nu G\frac{\partial{\cal L}_m}{\partial G}\,,
\ee
where
\be
K_{\mu\nu}\equiv \frac{\delta K}{\delta g^{\mu\nu}}=F_{\mu\rho}{F^\rho}_\nu\,,\label{K-tensor}
\ee
and its trace is given by\footnote{{Let us emphasise that the reader should pay attention to the fact that for the electromagnetic field $g^{\mu\nu}K_{\mu\nu}=2 K$ instead of $g^{\mu\nu}K_{\mu\nu}= K$.}}
\be
g^{\mu\nu}K_{\mu\nu} = {F^\mu}_\nu{F^\nu}_\mu = -F = 2K\,.
\ee
Using this notation, the energy-momentum tensor \eqref{T0} can be written as
\be
{T^\mu}_\nu=\left({\cal L}_m - G\frac{\partial{\cal L}_m}{\partial G}\right)\delta^\mu{}_\nu-2\frac{\partial{\cal L}_m}{\partial K}K^{\mu}{}_\nu\,.\label{T}
\ee
Given this result, the tensor structures that appear in the expansion (\ref{Om-T-Exp}) of the deformation matrix are $\delta^\mu{}_\nu$ and powers of $K^{\mu}{}_\nu$. The setting is very closely related to the case where the matter sector is described by scalar fields \cite{Afonso:2018hyj}, where arbitrary powers of $K^{\mu}{}_\nu$ turn out to be linear combinations of the identity and $K^{\mu}{}_\nu$ so that the correspondence between RBGs and GR can be explicitly unveiled. In fact, as shown in the appendix \ref{EMApp}, the following decomposition is always possible:
\be
K^\mu{}_{\lambda_1}K^{\lambda_1}{}_{\lambda_2}\cdots K^{\lambda_{p-1}}{}_\nu=a_p(G,K)\delta^\mu_\nu + b_p(G,K)K^\mu{}_\nu\,.\label{K-Product}
\ee
Therefore, the fundamental property that led to the explicit form of the correspondence between RBGs and GR for a scalar field matter sector, also holds for electromagnetic fields.
In this case, using the expansions (\ref{K-Product}) in (\ref{Om-T-Exp}), one readily sees  that the expansion \eqref{Om-T-Exp} leads to the following form for the deformation matrix
\be
\Omega^\mu{}_\nu=A(K,G)\delta^\mu_\nu+B(K,G)K^\mu{}_\nu\,,\label{DefMatExp}
\ee
with inverse
\be
\left(\Omega^{-1}\right)^\mu{}_\nu=C(K,G)\delta^\mu_\nu+D(K,G)K^\mu{}_\nu\,,\label{InvDefMatExp}
\ee
where the relation between the coefficients is
\be
A=\frac{C+DK}{C^2-D^2G^2+C D K}
 \qquad \ , \qquad
B=-\frac{D}{C^2-D^2G^2+C D K} \ ,\label{AB}
\ee
or equivalently
\be
C=\frac{A+BK}{A^2-B^2G^2+A B K}
 \qquad \ , \qquad
D=-\frac{B}{A^2-B^2G^2+A B K}\,.\label{CD}
\ee
Note that we have omitted the functional dependence of $A,B,C,D$ to lighten notation, and that the pairs of coefficients $(A,B)$ or $(C,D)$ are completely specified by (\ref{InvOm}) once a particular Gravity Lagrangian is found. A fundamental property of electromagnetic fields is closure under multiplication of the K-tensors defined in \eqref{K-tensor}. In fact,  \eqref{4FRed} can be rewritten in terms of the K-tensor as follows
\be
K^\mu{}_\rho K^\rho{}_\nu= G^2 \delta^\mu_\nu + K K^\mu{}_\nu\,.\label{KK}
\ee
Using this result, the determinant of the deformation matrix turns out to be
\be
\Omega = \left(A^2 - B^2 G^2 + A B K\right)^2 = \frac{1}{\left(C^2 - D^2 G^2 + C D K\right)^2}\,,\label{DetOmega}
\ee
which represents a key element to establish the correspondence between different frame invariants.

\subsection{Electromagnetic invariants in two frames}

In order to work out the correspondence between the original frame, represented by matter fields minimally coupled to $g_{\mu\nu}$, and the Einstein frame, where matter is minimally coupled to $q_{\mu\nu}$, a series of manipulations involving the deformation matrix $\hat\Omega$ are necessary. Introducing $\tilde F^\mu{}_\nu=q^{\mu\rho}F_{\rho\nu}=\left(\Omega^{-1}\right)^\mu{}_\sigma g^{\sigma\rho}F_{\rho\nu}$, where a tilde is used to denote objects whose indices are raised with $q^{\mu\nu}$ instead of $g^{\mu\nu}$, and using \eqref{BasicRel}, it is straightforward to prove that
\be
\tilde F^\mu{}_\nu=q^{\mu\rho}F_{\rho\nu}=\left(\Omega^{-1}\right)^\mu{}_\rho F^\rho{}_\nu = \left(C + D K\right)F^\mu{}_\nu - D\,G\,{}^\star F^\mu{}_\nu\,,
\ee
\be
{}^\star \tilde F^\mu{}_\nu = {}^\star \tilde F^{\mu\rho}q_{\rho\nu}=\Omega^{-1/2}\,\,g_{\nu\rho}\Omega^\rho{}_\sigma {}^\star  F^{\mu\sigma} =  \left(C + DK\right) {}^\star F^\mu{}_\nu + D G F^\mu{}_\nu\,.
\ee
The above relations can be written in the compact form
\begin{equation}
\left(\begin{array}{c} \tilde F^\mu{}_\nu \\ {}^\star \tilde F^\mu{}_\nu \end{array}\right) = \left(\begin{array}{cc}  \left(C + D K\right) & - D\,G \\ D G& ( C + D K) \end{array}\right)\left(\begin{array}{c}  F^\mu{}_\nu \\ {}^\star F^\mu{}_\nu \end{array}\right) \; ,
\end{equation}
that reminds of a duality rotation. Furthermore, from these equations it follows that
\begin{align*}
&\tilde K^\mu{}_\nu=D G^2 [2 C+D K] \delta^\mu{}_\nu+\left[(C+DK)^2+(DG)^2\right]K^\mu{}_\nu\\
&\tilde G^\mu{}_\nu = -G \left(C^2 - D^2 G^2 + C D K\right)\delta^\mu_\nu = -
\frac{G}{\sqrt{\Omega}}\delta^\mu_\nu \,
\end{align*}
where $\tilde K^\mu{}_\nu\equiv\tilde F^\mu{}_\rho\tilde F^\rho{}_\nu$ and $\tilde G^\mu{}_\nu\equiv\tilde F^\mu{}_\rho{}^\star\tilde F^\rho{}_\nu$  have been introduced. Tracing these equations, one finds the general relation between the electromagnetic invariants minimally coupled to $g_{\mu\nu}$ and $q_{\mu\nu}$, namely,
\begin{align}
&\tilde K= \left[(C+DK)^2+3(DG)^2\right] K+ 4 C D G^2\label{KinCorr}\\
&\tilde G = G \left(C^2 - D^2 G^2 + C D K\right) = G \Omega^{-1/2}\label{GCorr}
\end{align}
This proves that it is always possible to express the Einstein frame invariants $\tilde K$ and $\tilde{G}$ in terms of  the original invariants $K$ and $G$ of the RBG frame. The inverse relations, namely $K$ and $G$ in terms of $\tilde K$ and $\tilde{G}$, can be obtained in the same way noting that (following similar manipulations and notation)
\begin{equation}
\left(\begin{array}{c}  F^\mu{}_\nu \\ {}^\star F^\mu{}_\nu \end{array}\right) = \left(\begin{array}{cc}  \left(\tilde A + \tilde B \tilde K\right) & - \tilde B\,\tilde G \\ \tilde B\,\tilde G &  \left(\tilde A + \tilde B \tilde K\right) \end{array}\right)\left(\begin{array}{c} \tilde F^\mu{}_\nu \\ {}^\star \tilde F^\mu{}_\nu \end{array}\right) \ ,
\end{equation}
which leads to
\begin{align*}
&K^\mu{}_\nu=  \tilde B \tilde G^2 [2 \tilde A+\tilde B \tilde K] \delta^\mu{}_\nu+\left[(\tilde A+\tilde B \tilde K)^2+(\tilde B \tilde G)^2\right]\tilde K^\mu{}_\nu\,,\\
&G^\mu{}_\nu = -\tilde G \left[\tilde A\left(\tilde A +\tilde B\tilde K\right) - \tilde B^2 \tilde G^2\right] \delta^\mu{}_\nu\, .
\end{align*}
Taking the traces of these quantities, one finally obtains
\begin{align}
&K= \left[(\tilde A+\tilde B \tilde K)^2+3(\tilde B \tilde G)^2\right] \tilde K+ 4 \tilde A \tilde B \tilde G^2\,,\label{InvKinCorr}\\
&G= \tilde G \left[ \tilde A (\tilde A +  \tilde B \tilde K) - \tilde B^2 \tilde G^2 \right]=\tilde G \Omega^{1/2}\,,\label{InvGCorr}
\end{align}
which shows that if the $\hat\Omega$ matrix is known in terms of the Einstein frame variables, the corresponding electromagnetic invariants in the RBG frame can be obtained in terms of those in the Einstein frame.

\subsection{Correspondence between frames}

The correspondence suggested in (\ref{mapping}) combined with the field equations of the NED fields in the RBG and GR frames puts forward that the necessary relation between matter Lagrangians to map one frame into the other arises by simply demanding that the combination of the terms proportional to the identity in (\ref{mapping}) vanishes together with the combination of the other two non-diagonal terms. A detailed derivation and discussion of this point has been omitted here simply to lighten the technical part, as it does not provide any essential new insight (the interested reader is referred to Appendix \ref{App:EOM-NED}). After taking into account these considerations, the Einstein frame NED Lagrangian $\tilde {\cal L}_m(q_{\mu\nu},A_\alpha)$ associated to a given Lagrangian ${\cal L}_m(g_{\mu\nu},A_\alpha)$ can be written as
\be
\tilde {\cal L}_m = \Omega^{-1/2}\left[2\left(K\frac{\partial {\cal L}_m}{\partial K}+G\frac{\partial {\cal L}_m}{\partial G}\right)-{\cal L}_G-{\cal L}_m \right] \ , \label{eq:Map1}
\ee
which provides a parametric representation in terms of the $K$ and $G$ invariants of the RBG frame. The Lagrangian $\tilde {\cal L}_m$ can, in principle, be written as a function of its natural invariants $\tilde K$ and $\tilde G$ by inverting the relations (\ref{KinCorr}) and (\ref{GCorr}) to obtain $K(\tilde K,\tilde G)$ and  $G(\tilde K,\tilde G)$. This is only possible when a particular gravity Lagrangian is specified, which allows to explicitly construct the functions $C$ and $D$ that define the deformation matrix $\hat \Omega$. The nondiagonal part of (\ref{mapping}) leads to a relation between the partial derivatives of the matter Lagrangians, which takes the form
\be
\tilde K^\mu{}_\nu\frac{\partial \tilde {\cal L}_m}{\partial \tilde K} + \frac12 \delta^\mu_\nu\tilde G\frac{\partial \tilde {\cal L}_m}{\partial \tilde G}= \Omega^{-1/2}\,\left(K^\mu{}_\nu\frac{\partial {\cal L}_m}{\partial K}+\frac12 \delta^\mu_\nu G\frac{\partial {\cal L}_m}{\partial G}\right) \,.\label{eq:Map2}
\ee
By tracing this equation and using the result in (\ref{eq:Map1}), one may also find an expression for the RBG frame matter Lagrangian if the Einstein frame counterpart is provided, namely,
\be
{\cal L}_m = -{\cal L}_G + \Omega^{1/2}\left[2\left(\tilde K\frac{\delta \tilde{\cal L}_m}{\delta \tilde K} +\tilde G\frac{\delta \tilde{\cal L}_m}{\delta \tilde G}\right) - \tilde {\cal L}_m \right]\,,\label{eq:Map3}
\ee
which represents a parametrization of ${\cal L}_m$ in terms of the invariants of $\tilde K$ and $\tilde G$. Analogously to the previous case, using the relations (\ref{InvKinCorr}) and (\ref{InvGCorr}), one can obtain $\tilde K(K,G)$ and  $\tilde G(K,G)$ such that ${\cal L}_m$ is written in terms of its natural variables $K$ and $G$. Finally, an interesting relation between the modified gravity Lagrangian and the Legendre transform of the matter sectors, can be found combining  \eqref{eq:Map1}, \eqref{eq:Map2} and \eqref{eq:Map3} :
\be
L\left[{\cal L}_m\right] + \Omega^{1/2}L\left[\tilde{\cal L}_m\right] = {\cal L}_G\,,
\ee
where
\be
L\left[{\cal L}_m\right] = K\frac{\partial {\cal L}_m}{\partial K}+G\frac{\partial {\cal L}_m}{\partial G}-{\cal L}_m\,.
\ee
In terms of the traces of the energy momentum tensor in the two frames, this equation can be translated into
\be
T + \Omega^{1/2}\tilde T = {\cal L}_G\,,
\ee
where it is worth to remind that this equation holds on-shell.
Explicit examples will be worked out in the next sections.

\section{f(R) gravity theories Coupled to EM Field}

The gravitational sector of the action describing an $f(R)$ theory is given by
\begin{equation}\label{BIgrav}
\mathcal{S}_{f(R)} = \int{d^4 x\, \sqrt{-g} {\cal L}_G} = \frac{1}{2\kappa^2 } \int d^4 x \sqrt{-g} f ( R ) \ .
\end{equation}
Particularizing \eqref{q} for this subclass of RBG theories we obtain\be
\sqrt{|q|} q^{\mu\nu} = \sqrt{-g}\,f_R(R) g^{\mu\nu}\,,
\ee
leading to a conformal deformation matrix and the well known conformal relation between both metrics \cite{Olmo:2011uz}

\be
\Omega^{\mu}{}_{\nu} = f_R(R)\, \delta^{\mu}{}_{\nu}\,.
\qquad\text{and}\qquad q_{\mu\nu} = f_R(R) g_{\mu\nu}\,,\label{Metric-Rel}
\ee
where $f(R)$ has to be considered as an on-shell function of the (trace of the) matter stress-energy tensor. In order to find this dependence, one observes that the Ricci scalar, as a function of matter fields, can be obtained by solving the algebraic equation provided by the trace of the metric field equation \eqref{MetricfieldEq} when particularized to the $f(R)$ subclass:
\be
f_R(R)  R - 2 f(R) = \kappa^2 T\,,\label{Alg-Eq}
\ee
The matter action in the GR frame of $f(R)$ theories is thus provided by \eqref{eq:Map1} :
\be
\tilde {\cal L}_m = f_R^{-2}\left[2\left(K\frac{\partial {\cal L}_m}{\partial K}+G\frac{\partial {\cal L}_m}{\partial G}\right)-{\cal L}_m-\frac{1}{2\kappa}f \right]\label{f-par}\ ,
\ee
where $K$ and $G$ are the electromagnetic invariants in the RBG frame. In order to express these invariants in the Eoinstein frame, notice that the expansion \eqref{InvDefMatExp} is trivial for the $f(R)$ subclass, since there is only one non-vanishing coefficient $C=1/f_R$. This leads to the following relations between electromagnetic invariants in the two frames
\be
K = f_R \tilde K\,,\qquad G = f_R^2 \tilde G\,.\label{Inv-Rel},
\ee
which allows us to find the corresponding electromagnetic invarinats in the GR frame from \eqref{f-par}. Concerning the mapping from GR to the RBG frame within the $f(R)$ subclass, if the matter sector of the GR frame is given, then \eqref{eq:Map3} provides the recipe to find the matter sector in the RBG frame:
\be
{\cal L}_m = -\frac{1}{2\kappa}f + f_R^{2}\left[2\left(\tilde K\frac{\delta \tilde{\cal L}_m}{\delta \tilde K} +\tilde G\frac{\delta \tilde{\cal L}_m}{\delta \tilde G}\right) - \tilde {\cal L}_m \right]\,,\label{f-inv-par}
\ee
where
\be
\tilde K = f_R^{-1} K\,,\qquad \tilde G = f_R^{-2} G\,.
\ee
Let us now illustrate the above discussion with a simple example of a UV $f(R)$ correction. Consider thus the Starobinsky model described by the action
\be
S_s=\frac{1}{2\kappa^2}\int d^4 x\sqrt{-g}\left(R+\alpha R^2\right)\,,
\ee
where $\alpha=(6M^2)^{-1}$ has dimension of the inverse of length squared in the SI. In this case $f(R)=R+\alpha R^2$ so that  \eqref{Alg-Eq} implies that the curvature is proportional to the Legendre transform of the matter sector with respect to the electromagnetic invariants:
\be
R=-8\kappa^2\left({\cal L}_m - G\frac{\partial{\cal L}_m}{\partial G} -K\frac{\partial{\cal L}_m}{\partial K}\right)\,,\label{Curv-f(R)}
\ee
where we used the fact that the trace of the energy-momentum tensor is
\be
T = 4\left({\cal L}_m - G\frac{\partial{\cal L}_m}{\partial G} -K\frac{\partial{\cal L}_m}{\partial K}\right)\,.\label{T-Trace}
\ee
Using the above results, the relations between the electromagnetic invariants in different frames obtained by using \eqref{Inv-Rel} read
\be
\tilde K = \frac{K}{1-16\alpha\kappa^2\left({\cal L}_m - G\frac{\partial{\cal L}_m}{\partial G} -K\frac{\partial{\cal L}_m}{\partial K}\right)}\,,\qquad \tilde G = \frac{G}{\left[1-16\alpha\kappa^2\left({\cal L}_m - G\frac{\partial{\cal L}_m}{\partial G} -K\frac{\partial{\cal L}_m}{\partial K}\right)\right]^2}\,.
\ee
Given $G=G(\tilde G)$ and $F=F(\tilde F)$, the curvature \eqref{Curv-f(R)} can be expressed in terms of the GR frame and the matter Lagrangian \eqref{f-par} can be unveiled. Notice that the case in which the RBG frame matter sector is given by Maxwell electrodynamics (i.e. $L=K$) is too trivial due to the fact that the traceless energy-momentum tensor associated to the electromagnetic field necessarily implies zero curvature as a consequence of \eqref{Alg-Eq}. In turn, the metric field equations has the same form in both frames, namely
\be
R_{\mu\nu}=\kappa^2T_{\mu\nu}\,.
\ee
This implies that the space of solutions is the same, consistent with the fact that, due to the vanishing curvatures, it turns out that $f(T)=0$ and $f_R(T)=1$ so that from  \eqref{f-par}
\be
\tilde {\cal L}_m = K\ ,
\ee
and, according to \eqref{Metric-Rel}, the metrics are the same $q_{\mu\nu}=g_{\mu\nu}$.
Therefore we arrived to the well-known result that metric-affine Starobinsky theory gravity is equivalent to GR when both theories are minimally coupled to a free Maxwell electromagnetic field.

 Rather than exploring other nonlinear electromagnetic theories coupled to $f(R)$ theories, we will move on to consider a more appealing RBG model which in the weak field expansion recovers the Starobinsky theory plus additional quadratic and higher order corrections in the Ricci tensor, namely, the so-called EiBI gravity theory.

\section{Eddington-inspired Born-Infeld gravity theory Coupled to EM Fields}

We are now going to show how one can construct the Einstein frame matter Lagrangian $\tilde{\mathcal{L}}_m$ that is associated with a matter Lagrangian $\mathcal{L}_m$ coupled to the so-called Eddington-inspired Born-Infeld gravity (EiBI). Vice versa, we will also obtain the form of the Lagrangian  $\mathcal{L}_m$  when one starts with GR coupled to $\tilde{\mathcal{L}}_m$. We begin by defining the action of the EiBI theory as a UV (or high-curvature) modification of GR of the form
\begin{equation}\label{BIgrav}
\mathcal{S}_{EiBI}=\int{d^4 x\, \sqrt{-g} {\cal L}_G} = \frac{1}{\kappa^2 \epsilon} \int d^4 x \left[\sqrt{-\left|g_{\mu\nu}+\epsilon R_{\mu\nu}\right|}-\lambda \sqrt{-g}\right] \ .
\end{equation}
For a recent review on this theory and associated models see \cite{BeltranJimenez:2017doy}. The deformation matrix within EiBI as given by \eqref{Om} reads
\be
\Omega^\mu{}_\nu = \delta^\mu{}_\nu +\epsilon R^\mu{}_\nu\,,
\ee
which leads to an auxiliary metric 
\be
q_{\mu\nu} = g_{\mu\nu}+\epsilon R_{\mu\nu}\,.
\ee
As well known, we can express the EiBI lagrangian in terms of the determinant of the deformation matrix as
\be
{\cal L}_G = \frac{\Omega^{\frac12}-\lambda}{\epsilon\kappa^2}\,,
\ee
and the EiBI metric field equations obtained from \eqref{PreEinstein} read
\be
2\kappa^2\frac{\partial{\cal L}_G}{\partial R_{\mu\rho}} g_{\rho\nu} = \lambda\delta^\mu_\nu-\epsilon\kappa^2 T^\mu{}_\nu\,.
\ee
which using \eqref{OmG} leads to an on-shell relation between the deformation matrix and the stress-energy tensor of the matter sector of the form
\be
\sqrt{\Omega}\left(\Omega^{-1}\right)^\mu{}_\nu = \lambda\delta^\mu_\nu-\epsilon\kappa^2 T^\mu{}_\nu\,.\label{Om-def}
\ee

From this relation, and using the stress-energy tensor for a general electrodynamics, one can find the explicit expression of the determinant of the deformation matrix in EiBI coupled to a general electrodynamics, given by
\be
\Omega = \left\{\left[\lambda + \epsilon \kappa^2\left(G\frac{\partial{\cal L}_m}{\partial G}-
{\cal L}_m\right)\right]^2 - \left( 2 \epsilon\kappa^2 \frac{\partial{\cal L}_m}{\partial K}\right)^2 G^2 + 2 \epsilon\kappa^2 K\left[\lambda + \epsilon \kappa^2\left(G\frac{\partial{\cal L}_m}{\partial G}-
{\cal L}_m\right)\right]\frac{\partial{\cal L}_m}{\partial K}\right\}^2\,.\label{DetOmEiBI}
\ee

where \eqref{T} has been used. From \eqref{Om-def}, it is straightforward to find that the expansion \eqref{InvDefMatExp} leads to a form of the C and D coefficients given by
\be
C=\Omega^{-1/2}\left[\lambda + \epsilon \kappa^2\left(G\frac{\partial{\cal L}_m}{\partial G}-
{\cal L}_m\right)\right]\,,\qquad
D=2\epsilon \kappa^2\Omega^{-1/2} \frac{\partial{\cal L}_m}{\partial K}\,,
\ee

The above expressions and manipulations clarify how to obtain the deformation matrix, its determinant, and the various coefficients that were formally used in our general manipulations in Section \ref{sec:RBG_NEDs}. We thus have all the ingredientes to proceed and construct the Einstein-frame matter Lagrangian associated with the original $\mathcal{ L}_m$. The equations describing the correspondence \eqref{eq:Map1} and \eqref{eq:Map2} read
\be
\tilde{\cal L}_m=\Omega^{-\frac12}\left[2\left(K\frac{\partial{\cal L}_m}{\partial K}+G\frac{\partial{\cal L}_m}{\partial G}\right)-{\cal L}_m+\frac{\lambda}{\epsilon\kappa^2}\right]-\frac{1}{\epsilon\kappa^2}\,,\label{CorrEiBIGR1}
\ee
\be
\tilde K^\mu{}_\nu\frac{\partial \tilde {\cal L}_m}{\partial \tilde K} + \frac12 \delta^\mu_\nu\tilde G\frac{\partial \tilde {\cal L}_m}{\partial \tilde G}= \Omega^{-1/2}\,\left(K^\mu{}_\nu\frac{\partial {\cal L}_m}{\partial K}+\frac12 \delta^\mu_\nu G\frac{\partial {\cal L}_m}{\partial G}\right)\, .\label{CorrEiBIGR2}
\ee
Inserting now our previous results in (\ref{CorrEiBIGR1}), we get 
\be
\tilde{\cal L}_m=\frac{2\left(K\frac{\partial{\cal L}_m}{\partial K}+G\frac{\partial{\cal L}_m}{\partial G}\right)-{\cal L}_m+\frac{\lambda}{\epsilon\kappa^2}}{\left[\lambda + \epsilon \kappa^2\left(G\frac{\partial{\cal L}_m}{\partial G}-
{\cal L}_m\right)\right]^2 - \left( 2 \epsilon\kappa^2 \frac{\partial{\cal L}_m}{\partial K}\right)^2 G^2 + 2 \epsilon\kappa^2 K\left[\lambda + \epsilon \kappa^2\left(G\frac{\partial{\cal L}_m}{\partial G}-
{\cal L}_m\right)\right]\frac{\partial{\cal L}_m}{\partial K}}
-\frac{1}{\epsilon\kappa^2} \,.\label{TildeL}
\ee
which provides a parametric representation of the Einstein-frame matter Lagrangian $\tilde{\cal L}_m$ in terms of the original RBG frame invariants $K$ and $G$. We are now going to show how one can deal with the inverse problem, namely, go from the Einstein frame matter variables to the RBG frame. To that end, one can use  \eqref{CorrEiBIGR1} and \eqref{CorrEiBIGR2} to obtain the expansion \eqref{InvDefMatExp} 
\be
\left(\Omega^{-1}\right)^\mu{}_\nu=\tilde C\delta^\mu_\nu+\tilde D \tilde K^\mu{}_\nu\,,
\ee

in Einstein frame variables. Given that in general 
\bea
\left(\Omega^{-1}\right)^\mu{}_\nu =\left[1 - \epsilon \kappa^2\left(\tilde K\frac{\partial\tilde{\cal L}_m}{\partial \tilde K}+\tilde G\frac{\partial\tilde{\cal L}_m}{\partial \tilde G}-
\tilde{\cal L}_m\right)\right]\delta^\mu_\nu + 2\epsilon \kappa^2\frac{\partial\tilde{\cal L}_m}{\partial \tilde K}\tilde K^\mu{}_\nu,
\eea
we find the relations
\be
\tilde C=1 - \epsilon \kappa^2\left(\tilde K\frac{\partial\tilde{\cal L}_m}{\partial \tilde K}+\tilde G\frac{\partial\tilde{\cal L}_m}{\partial \tilde G}-
\tilde{\cal L}_m\right)\,,\qquad \tilde D = 2\epsilon\kappa^2\frac{\partial\tilde{\cal L}_m}{\partial \tilde K}\,,
\ee
which lead to the following expression for the determinant of the deformation matrix in the Einstein frame variables:
\bea
\Omega = \left\{\left[1 - \epsilon \kappa^2\left(\tilde K\frac{\partial\tilde{\cal L}_m}{\partial \tilde K}+\tilde G\frac{\partial\tilde{\cal L}_m}{\partial \tilde G}-
\tilde{\cal L}_m\right)\right]^2 - 4\left(\epsilon\kappa^2\tilde G\frac{\partial\tilde{\cal L}_m}{\partial \tilde K}\right)^2 +\right.\nonumber\\
+ \left. 2\epsilon\kappa^2\tilde K\frac{\partial\tilde{\cal L}_m}{\partial \tilde K}\left[1 - \epsilon \kappa^2\left(\tilde K\frac{\partial\tilde{\cal L}_m}{\partial \tilde K}+\tilde G\frac{\partial\tilde{\cal L}_m}{\partial \tilde G}-
\tilde{\cal L}_m\right)\right]
\right\}^{-1}\,.\label{DetOmGR}
\eea
Using the above relations for EIBI coupled to a generic electrodynamics, the RBG frame matter Lagrangian can thus be written in terms of the Einstein frame matter fields as
\bea
{\cal L}_m =\frac{
2\left(\tilde K\frac{\partial{\cal \tilde L}_m}{\partial \tilde K} +\tilde G\frac{\partial{\cal \tilde L}_m}{\partial \tilde G}\right) - \tilde{\cal L}_m - \frac{1}{\epsilon\kappa^2}
}
{
\left[
1 - \epsilon \kappa^2\left(\tilde K\frac{\partial\tilde{\cal L}_m}{\partial \tilde K}+\tilde G\frac{\partial\tilde{\cal L}_m}{\partial \tilde G}-\tilde{\cal L}_m\right)
\right]^2
- 4\left(\epsilon\kappa^2\tilde G \frac{\partial\tilde{\cal L}_m}{\partial \tilde K}
\right)^2 + 2\epsilon\kappa^2\tilde K\frac{\partial\tilde{\cal L}_m}{\partial \tilde K}\left[1 - \epsilon \kappa^2\left(\tilde K\frac{\partial\tilde{\cal L}_m}{\partial \tilde K}+\tilde G\frac{\partial\tilde{\cal L}_m}{\partial \tilde G}-
\tilde{\cal L}_m\right)\right]
}
+
\frac{\lambda}{\epsilon\kappa^2} \ ,\label{EiBI-InvMap}
\eea
thus obtaining the desired result. We will next focus on a particular case of physical interest to show the capabilities of the mathematical machinery developed so far.

\subsection{Mapping Maxwell electromagnetism coupled to EiBI gravity into GR}

As a conservative starting point we will now assume modifications only in the gravitational sector, which we will assume to be described by EiBI theory, while the matter sector in its RBG frame is described by the standard Maxwell Lagrangian ${\cal L}_m = K$. For this theory, the Einstein frame Lagrangian readily follows from (\ref{TildeL}), leading to
\be\label{eq:Lraw}
\tilde{\cal L}_m = \frac{\lambda + \epsilon\kappa^2 K}{\epsilon\kappa^2\left[\lambda^2 - \epsilon^2\kappa^4 (K^2 + 4G^2)\right]} - \frac{1}{\epsilon\kappa^2}  \ .
\ee
This Einstein-frame Lagrangian is written in terms of the EiBI frame variables $(K,G)$. The next step is thus  to find the relation between the invariants $(K,G)$ and $(\tilde{K},\tilde{G})$ in order to be able to write $\tilde{\cal L}_m$ using its natural variables. To proceed in this direction, we need first to obtain the coefficients of the inverse deformation matrix (\ref{InvDefMatExp}), which take the form
\be
C=\frac{\lambda - \epsilon \kappa^2 K}{\lambda^2 -  \epsilon^2\kappa^4\left(K^2 +4G^2\right)}\,,\qquad
D=\frac{2\epsilon \kappa^2}{\lambda^2 -  \epsilon^2\kappa^4\left(K^2 +4G^2\right)} \,,
\ee
together with the determinant $\Omega = \left[\lambda^2 -  \epsilon^2\kappa^4\left(K^2 +4G^2\right)\right]^{2}$. Inserting this in (\ref{KinCorr}) and (\ref{GCorr}), the electromagnetic invariants become
\be
\tilde K = \frac{\lambda^2 K + \epsilon\kappa^2\left(2\lambda + \epsilon\kappa^2 K\right)\left(K^2 +4G^2\right)}{\left[\lambda^2 - \epsilon^2\kappa^4\left(K^2 +4G^2\right)\right]^2}
\ee
\be
\tilde G = \frac{ G}{\left[\lambda^2 - \epsilon^2\kappa^4\left(K^2 +4G^2\right)\right]}\,.
\ee
Despite appearances, these equations can be easily inverted to provide an expression of the invariants $K$ and $G$ in terms of the GR frame invariants $\tilde K$ and $\tilde G$, namely
\be
K = \frac{\left(\tilde K - 8\epsilon\kappa^2 \lambda\tilde G^2\right)\left(1+2\epsilon\kappa^2\lambda\tilde K \pm \sqrt{1+ 4\epsilon\kappa^2\lambda\tilde K-16\epsilon^2\kappa^4 \lambda^2 \tilde G^2}\right)}{2\epsilon^2\kappa^4\left(4\tilde G^2 + \tilde K^2\right)}\nonumber
\ee
\be
G = -\frac{\tilde G\left[1+4\epsilon\kappa^2\lambda\tilde K - 16\epsilon^2\kappa^4 \lambda^2 \tilde G^2   \pm \left(1+2\epsilon\kappa^2\lambda\tilde K\right) \sqrt{1+ 4\epsilon\kappa^2\lambda\tilde K-16\epsilon^2\kappa^4 \lambda^2 \tilde G^2}\right]}{2\epsilon^2\kappa^4\left(4\tilde G^2 + \tilde K^2\right)}\nonumber \ .
\ee
Now we just need to substitute these expressions into (\ref{eq:Lraw}) to obtain
\be
\tilde{\cal L}_m = \frac{1-2\lambda\pm \sqrt{1+ 4\epsilon\kappa^2\lambda(\tilde K-4\epsilon\kappa^2\lambda \tilde G^2)}}{2\epsilon\kappa^2\lambda}\, .
\ee
In the asymptotically flat case, $\lambda\to 1$, taking the positive sign in front of the square root, and redefining $4\epsilon\kappa^2\to -1/\beta^2$, this Lagrangian turns into
\be
\tilde {\cal L}_{BI} = 2\beta^2\left(1-\sqrt{1
-\frac{\tilde K}{\beta^2} - \frac{\tilde G^2}{\beta^4}}\right)= 2\beta^2\left(1-\sqrt{1
+\frac{1}{2\beta^2}F_{\mu\nu}\tilde{F}^{\mu\nu} - \frac{1}{16\beta^4}(F_{\mu\nu}{}^\star \tilde{F}^{\mu\nu})^2}\right)\ ,
\ee
which is nothing but the well-known Born-Infeld electromagnetic Lagrangian \cite{Born:1934gh}. This constitutes the most relevant practical result of this work. It shows that the correspondence between EiBI gravity coupled to Maxwell electromagnetism with GR coupled to Born-Infeld electromagnetism occurs not only in the static case, but in full generality. In particular, this suggests that information about the Born-Infeld gravitational sector in the RBG frame can be derived from the properties of the Born-Infeld electromagnetic theory in GR, which establishes a deep relation between the geometric and matter sectors of these theories. The physical implications of this correspondence will be explored in forthcoming works.

\section{Summary and Conclusions} \label{sec:Summary}

In this work we provided a general method to relate arbitrary  RBG theories coupled to arbitrary electrodynamics with GR coupled to other electrodynamics. This relation has been studied both at the level of field equations as at the Lagrangian level. To make this approach more concrete, specific examples of RBGs have been considered, namely a quadratic $f(R)$ model and the so-called Eddington-inspired Born-Infeld theory of gravity, both involving higher curvature corrections to Einstein's theory. In order to establish a correspondence that is independent of the specific metric, it has been crucial to recast the equations in terms of the fundamental electromagnetic invariants $K=-\frac{1}{2}F_{\mu\nu}F^{\mu\nu}$ and $G=\frac{1}{4}F_{\mu\nu}\star F^{\mu\nu}$. A key step has been the decomposition of the electromagnetic stress-energy tensor in terms of the invariants whose algebraic properties allow to express arbitrary powers of the tensor $K^\mu{}_\nu\equiv F^{\mu\alpha}F_{\alpha\nu}$ in terms of a linear combination of itself and the identity (see Appendix \ref{EMApp}). In addition, consistency of the correspondence between the field equations of the electromagnetic field in the two frames, have been used to obtain simplified relations that unveil the electrodynamics of a frame once the matter sector of the other frame is given, as detailed in Appendix \ref{App:EOM-NED}. 

The most important practical result of this paper is that we have shown that the EiBI gravity model coupled to Maxwell electromagnetism turns out to be in exact correspondence with GR coupled to Born-Infeld electromagnetism. This makes it explicit that the matter and gravity sectors are intimately related in a deep and {\it a priori} unexpected way. As a consequence, the propagation of Maxwell electromagnetic waves in the EiBI theory is fully determined by how Born-Infeld electromagnetic waves propagate under the dynamics of GR. In this sense, one finds that Born-Infeld electromagnetic waves in GR do not follow geodesics of the background metric due to the nonlinear character of their evolution equations. Those trajectories, however, would correspond to null geodesics of the EiBI gravity metric. Therefore, one expects an apparent degeneracy between the effects that nonlinearities in the matter sector would cause in GR and those caused on linear matter Lagrangians in RBGs. An indepth analysis of such phenomena and their phenomenological implications is currently underway to determine if the modified dynamics induced by nonlinearities of the matter sector can be distinguished from those induced by the gravitational sector on linear field Lagrangians.

\section*{Acknowledgments}
G.J.O. thank Universidade Federal do Rio Grande do Norte and the International Institute of Physics of Natal for kind hospitality. A. D. is supported by a PhD contract of the program FPU 2015 (Spanish Ministry of Economy and Competitiveness) with reference FPU15/05406.
G. J. O. is funded by the Ramon y Cajal contract RYC-2013-13019 (Spain).  This work is supported by the Spanish projects FIS2014-57387-C3-1-P  and FIS2017-84440-C2-1-P (MINECO/FEDER, EU), the project SEJI/2017/042 (Generalitat Valenciana), the Consolider Program CPANPHY-1205388, and the Severo Ochoa grant SEV-2014-0398 (Spain).
The authors would like to acknowledge partial financial support from the European Union's Horizon 2020 research and innovation programme under the H2020-MSCA-RISE-2017 Grant No. FunFiCO-777740.

\appendix

\section{Basic Properties of EM Field\label{EMApp}}

Introducing the generalized Kronecker delta as 
\begin{equation}
\delta ^{i_1 \cdots i_k}_{j_1 \cdots j_k} \equiv \frac{1}{k!} \det
\left(
        \begin{array}{cccc}
         \delta^{i_1}_{j_1} & \delta^{i_2}_{j_1} & \cdots & \delta^{i_k}_{j_1} \\
         \delta^{i_1}_{j_{2}} & \delta^{i_{2}}_{j_{2}} & \cdots & \delta^{i_k}_{j_{2}} \\
         \vdots & \vdots & \ddots & \vdots \\
         \delta^{i_1}_{j_k} & \delta^{i_{2}}_{j_k} & \cdots & \delta^{i_k}_{j_k} \\
        \end{array}
        \right) =\delta^{i_1}_{[j_1}\cdots \delta^{i_k}_{j_k]}\,,\label{Delta:k}
\end{equation}
one can obtain the following expression for the product of rank-4 Levi-Civita symbols
\begin{equation}
\epsilon^{\mu\nu\rho\sigma}\epsilon_{\mu\alpha\beta\gamma}=-3!\delta^{\nu\rho\sigma}_{\alpha\beta\gamma}\,,\quad \epsilon^{\mu\nu\rho\sigma}\epsilon_{\mu\nu\alpha\beta}=-2!2!\delta^{\rho\sigma}_{\alpha\beta}\,.
\end{equation}
Using the usual definition of the Hodge dual in holomic coordinates, namely :
\begin{equation}
{}^\star F_{\mu\nu} = \frac{\sqrt{|g|}}{2} \epsilon_{\mu\nu\rho\sigma}F^{\rho\sigma}\,,\qquad F_{\mu\nu}=-\frac{1}{\sqrt{2|g|}}\epsilon_{\mu\nu\rho\sigma}{}^\star F^{\rho\sigma}
\end{equation}
one finds
\begin{equation}
{}^\star F^\mu{}_\lambda {}^\star F^\lambda{}_\nu=-\frac14 \epsilon^{\lambda\mu\rho\sigma}\epsilon_{\lambda\nu\alpha\beta}F^{\alpha\beta}F_{\rho\sigma}=\frac32 \delta^{\mu\rho\sigma}_{\nu\alpha\beta}F^{\alpha\beta}F_{\rho\sigma}=\frac32\frac16\left(2\delta^\mu_\nu F^2+4F^\mu_\lambda F^{\lambda}{}_\nu\right)=\frac12\delta^\mu_\nu F + F^\mu_\lambda F^{\lambda}{}_\nu
\end{equation}
\begin{equation}
{}^\star F^\mu{}_\lambda F^\lambda{}_\nu = -{}^\star F^{\lambda\mu} F_{\lambda\nu}=-\left(\frac12\epsilon^{\lambda\mu\rho\sigma}F_{\rho\sigma}\right)\left(-\frac12\epsilon_{\lambda\nu\alpha\beta}{}^\star F^{\alpha\beta}\right)=-\frac{3!}{4}\delta^{\mu\rho\sigma}_{\nu\alpha\beta}F_{\rho\sigma}{}^\star F^{\alpha\beta}=-2\delta^\mu_\nu G -^\star F^\mu{}_\lambda F^\lambda{}_\nu\,.
\end{equation}
where $F\equiv  F_{\alpha\beta}F^{\alpha\beta}$ and $G\equiv  \frac14 F_{\alpha\beta}{}^\star F^{\alpha\beta}$ have been introduced.
Therefore, the following relations hold
\begin{equation}
F^\mu{}_\lambda F^{\lambda}{}_\nu={}^\star F^\mu{}_\lambda {}^\star F^\lambda{}_\nu-\frac12\delta^\mu_\nu F
\,,\qquad {}^\star F^\mu{}_\lambda F^\lambda{}_\nu = F^\mu{}_\lambda {}^\star F^\lambda{}_\nu = - \delta^\mu_\nu G\,.\label{BasicRel}
\end{equation}
Using this equations, products of $p$ field stenght can be reduced as follows:
\begin{equation}
F^{\mu}{}_{\lambda_1}F^{\lambda_1}{}_{\lambda_2}\cdots F^{\lambda_{p-2}}{}_{\lambda_{p-1}} F^{\lambda_{p-1}}{}_{\lambda_p}=
	F^\mu{}_{\lambda_1}F^{\lambda_1}{}_{\lambda_2}\cdots F^{\lambda_{p-5}}{}_{\lambda_p}	
	G^2
	-\frac12
	F^\mu{}_{\lambda_1}F^{\lambda_1}{}_{\lambda_2}\cdots F^{\lambda_{p-3}}{}_{\lambda_{p-2}} F\,.
	\label{FundEquat}
\end{equation}
Iterating this reduction equation, it is possible to write any product of field strenghts in terms of just two tensorial structures, namely the identity and $F^\mu{}_\rho F^\rho{}_\nu$ according to
\begin{equation}
F^{\mu}{}_{\lambda_1}F^{\lambda_1}{}_{\lambda_2}\cdots F^{\lambda_{p-2}}{}_{\lambda_{p-1}} F^{\lambda_{p-1}}{}_{\lambda_p}=
	a_p{(F,G)}\delta^\mu_\nu + b_p{(F,G)}F^\mu{}_{\rho}F^{\rho}{}_{\nu}\,.
	\label{TotRed}
\end{equation}
In particular, the following identity holds
\begin{equation}
F^{\mu}{}_{\lambda_1}F^{\lambda_1}{}_{\lambda_2}F^{\lambda_{2}}{}_{\lambda_{3}} F^{\lambda_{3}}{}_{\nu}=
	G^2
	-\frac12
	F^\mu{}_{\lambda_1}F^{\lambda_1}{}_{\nu} F\,.
	\label{4FRed}
\end{equation}

\section{Consistency of the mapping with the NED field equations}\label{App:EOM-NED}

 \eqref{mapping} have to be consistent with the field equations associated to the matter fields in the two frames. Performing the variation of the matter action with respect to $A_\mu$, one obtains the following field equations in the RBG and GR frames, respectively,
\be
\partial_\mu\left[\sqrt{-g}\left(\frac{\partial {\cal L}_m}{\partial K}F^{\mu\nu} - \frac12 \frac{\partial {\cal L}_m}{\partial G}{}^\star F^{\mu\nu}\right)\right] + \frac12 \sqrt{-g}\frac{\partial {\cal L}_m}{\partial A_\nu}=0\label{FE1}
\ee
\be
\partial_\mu\left[\sqrt{-q}\left(\frac{\partial \tilde{\cal L}_m}{\partial \tilde K} \tilde F^{\mu\nu} - \frac12  \frac{\partial\tilde{\cal L}_m}{\partial \tilde G}{}^\star \tilde F^{\mu\nu}\right)\right] + \frac12 \sqrt{-q}\frac{\partial \tilde{\cal L}_m}{\partial A_\nu}=0\,.\label{FE2}
\ee
Subtracting \eqref{FE1} from \eqref{FE2}, one finds
\be
\partial_\mu\left[\sqrt{-g}\left(\frac{\partial {\cal L}_m}{\partial K}F^{\mu\nu} - \frac12 \frac{\partial {\cal L}_m}{\partial G}{}^\star F^{\mu\nu}\right) - \sqrt{-q}\left(\frac{\partial \tilde{\cal L}_m}{\partial \tilde K} \tilde F^{\mu\nu} - \frac12  \frac{\partial\tilde{\cal L}_m}{\partial \tilde G}{}^\star \tilde F^{\mu\nu}\right)\right] = \frac12 \sqrt{-q}\frac{\partial \tilde{\cal L}_m}{\partial \tilde A_\nu}-
\frac12 \sqrt{-g}\frac{\partial {\cal L}_m}{\partial A_\nu} \,.\label{FE3}
\ee
Both sides of this equation should vanish independently so that the dependence of the Lagrangians on the electromagnetic invariants $K, G$ and $\tilde{K}, \tilde{G}$ must satisfy
\be
\sqrt{-g}\left(\frac{\partial {\cal L}_m}{\partial K}F^{\mu\nu} - \frac12 \frac{\partial {\cal L}_m}{\partial G}{}^\star F^{\mu\nu}\right) - \sqrt{-q}\left(\frac{\partial \tilde{\cal L}_m}{\partial \tilde K} \tilde F^{\mu\nu} - \frac12  \frac{\partial\tilde{\cal L}_m}{\partial \tilde G}{}^\star \tilde F^{\mu\nu}\right) = \Lambda^{\mu\nu}\label{FundConsEq}
\ee
while the explicit dependence of the vector potential has to be such that
\be
\frac{\partial\tilde{\cal L}_m}{\partial  A_\mu} = \Omega^{-1/2}\frac{\partial {\cal L}_m}{\partial A_\mu} +\frac{1}{\sqrt{-q}}\partial_\mu\Lambda^{\mu\nu}\,.\label{ConsEq}
\ee
We can now rearrange  \eqref{mapping} as follows
\be
\delta^\mu_\nu\left(g^{\rho\sigma}\frac{\delta{\cal L}_m(g_{\mu\nu},\psi)}{\delta g^{\rho\sigma}}-\mathcal{L}_G-\mathcal{L}_m(g_{\mu\nu},\psi)-\sqrt{\Omega}\tilde{\mathcal{L}}_m({q}_{\mu\nu},\psi)\right)
=
2\kappa^2 \left( g^{\mu\rho}\frac{\delta{\cal L}_m(g_{\mu\nu},\psi)}{\delta g^{\rho\nu}} - \sqrt{\Omega}\,\,{q}^{\mu\rho}\frac{\delta\tilde{\cal L}_m({q}_{\mu\nu},\psi)}{\delta {q}^{\rho\nu}}\right)\label{RearrMapping}
\ee
and focus on the right-hand side of this equation which, once multiplied by a $\sqrt{-g}/(2\kappa^2)$ factor, can be expanded in terms of derivatives of $K$ and $G$ as follows
\begin{align}
&\left( \sqrt{-g} g^{\mu\rho}\frac{\delta{\cal L}_m(g_{\mu\nu},\psi)}{\delta g^{\rho\nu}} - \sqrt{-q}\,\,{q}^{\mu\rho}\frac{\delta\tilde{\cal L}_m({q}_{\mu\nu},\psi)}{\delta {q}^{\rho\nu}}\right) = \nonumber\\
&= \left[ \sqrt{-g}\left(\frac{\partial{\cal L}_m}{\partial K}F^{\mu\sigma} - \frac12\frac{\partial{\cal L}_m}{\partial G}{}^\star F^{\mu\sigma}\right) - \sqrt{-q}\left(\frac{\partial\tilde{\cal L}_m}{\partial \tilde K}\tilde F^{\mu\sigma} - \frac12\frac{\partial\tilde{\cal L}_m}{\partial \tilde G}{}^\star \tilde F^{\mu\sigma}\right)\right]F_{\sigma\nu}\,.
\end{align}
In light of the above expression, assuming compatibility of the matter field equations in the RBG and GR frames is equivalent to stating that the right-hand side  of \eqref{RearrMapping} is related to $\Lambda^{\mu\nu}$ as follows
\be
 \left( g^{\mu\rho}\frac{\delta{\cal L}_m(g_{\mu\nu},\psi)}{\delta g^{\rho\nu}} - \sqrt{\Omega}\,\,{q}^{\mu\rho}\frac{\delta\tilde{\cal L}_m({q}_{\mu\nu},\psi)}{\delta {q}^{\rho\nu}}\right)=\frac{1}{\sqrt{-g}}\Lambda^{\mu\rho}F_{\rho\nu} \ ,
\ee
which leads to the following condition
\begin{align}
\delta^\mu_\nu\left[2\left(K\frac{\partial {\cal L}_m}{\partial K}+G\frac{\partial {\cal L}_m}{\partial G}\right) - \mathcal{L}_G-\mathcal{L}_m(g_{\mu\nu},\psi)-\sqrt{\Omega}\tilde{\mathcal{L}}_m({q}_{\mu\nu},\psi)\right]=\frac{2\kappa^2}{\sqrt{-g}}\Lambda^{\mu\rho}F_{\rho\nu}\,.
\end{align}
On the other hand, multiplying  \eqref{FundConsEq} by $F_{\nu\rho}$ and rearranging terms, one finds
\be
\sqrt{-g}\left(K^\mu{}_\nu\frac{\partial {\cal L}_m}{\partial K}+\frac12 \delta^\mu_\nu G\frac{\partial {\cal L}_m}{\partial G}\right) - \sqrt{-q}\left(\,\tilde K^\mu{}_\nu\frac{\partial \tilde {\cal L}_m}{\partial \tilde K} + \frac12 \delta^\mu_\nu\tilde G\frac{\partial \tilde {\cal L}_m}{\partial \tilde G}\right) = \Lambda^{\mu\rho}F_{\rho\nu}\,.\label{FundConsEq2}
\ee
Given that the left-hand side of (\ref{FE3}) should vanish independently of its right-hand side, it seems natural to take $\Lambda^{\mu\nu}=0$. With this restriction, or simply assuming that $\Lambda^{\mu\rho}F_{\rho\nu}=0$, the equations describing the correspondence become
\be
\tilde {\cal L}_m = \Omega^{-1/2}\left[2\left(K\frac{\partial {\cal L}_m}{\partial K}+G\frac{\partial {\cal L}_m}{\partial G}\right)-{\cal L}_G-{\cal L}_m \right]\label{Map1}
\ee
\be
\tilde K^\mu{}_\nu\frac{\partial \tilde {\cal L}_m}{\partial \tilde K} + \frac12 \delta^\mu_\nu\tilde G\frac{\partial \tilde {\cal L}_m}{\partial \tilde G}= \Omega^{-1/2}\,\left(K^\mu{}_\nu\frac{\partial {\cal L}_m}{\partial K}+\frac12 \delta^\mu_\nu G\frac{\partial {\cal L}_m}{\partial G}\right) \,.\label{Map2}
\ee
The first of the above equations provides a parametric representation of the matter Lagrangian in the GR frame in terms of the invariants of the RBG frame. By contracting indices in the second equation, one finds a relation between the partial derivatives of the matter Lagrangians with respect to their arguments. Note also that the traced equation,
\be
\tilde K\frac{\partial \tilde {\cal L}_m}{\partial \tilde K} - \tilde G\frac{\partial \tilde {\cal L}_m}{\partial \tilde G}= \Omega^{-1/2}\,\left(K\frac{\partial {\cal L}_m}{\partial K} - G\frac{\partial {\cal L}_m}{\partial G}\right) \ ,\label{TraceMap2}
\ee
can be substituted in \eqref{Map1} to find
\be
{\cal L}_m = -{\cal L}_G + \Omega^{1/2}\left[2\left(\tilde K\frac{\delta \tilde{\cal L}_m}{\delta \tilde K} +\tilde G\frac{\delta \tilde{\cal L}_m}{\delta \tilde G}\right) - \tilde {\cal L}_m \right]\,,\label{Map3}
\ee
which represents a parametrization of the matter Lagrangian in the RBG frame in terms of the invariants of the GR frame whenever $\Omega$ and ${\cal L}_G$ can be expressed in terms of fields in the GR frame.



\begin{thebibliography}{100}

\bibitem{Latorre:2017uve} 
  A.~Delhom, G.~J.~Olmo and M.~Ronco,
  Phys.\ Lett.\ B {\bf 780}, 294 (2018)
  doi:10.1016/j.physletb.2018.03.002
  [arXiv:1709.04249 [hep-th]].
  
 \bibitem{PenuelasMirallesDelhom}
   A.~Delhom, V. Miralles and A. Pe\~nuelas,
   \textit{To appear (2019).}
  

\bibitem{Olmo:2019qsj} 
  G.~J.~Olmo, D.~Rubiera-Garcia and A.~Wojnar,
  arXiv:1906.04629 [GR qc].  
  
\bibitem{Afonso:2018bpv}
  V.~I.~Afonso, G.~J.~Olmo and D.~Rubiera-Garcia,
  Phys.\ Rev.\ D {\bf 97}, 021503 (2018).

\bibitem{Afonso:2018hyj}
  V.~I.~Afonso, G.~J.~Olmo, E.~Orazi and D.~Rubiera-Garcia,
  Phys.\ Rev.\ D {\bf 99}, no. 4, 044040 (2019)
  [arXiv:1810.04239 [GR qc]].
  

\bibitem{Afonso:2019fzv} 
  V.~I.~Afonso, G.~J.~Olmo, E.~Orazi and D.~Rubiera-Garcia,
  arXiv:1906.04623 [hep-th].
  
\bibitem{Afonso:2018mxn}
  V.~I.~Afonso, G.~J.~Olmo, E.~Orazi and D.~Rubiera-Garcia,
  Eur.\ Phys.\ J.\ C {\bf 78}, no. 10, 866 (2018)
  doi:10.1140/epjc/s10052-018-6356-1
  [arXiv:1807.06385 [GR qc]].
  
\bibitem{Afonso:2017bxr}
  V.~I.~Afonso, C.~Bejarano, J.~Beltran Jimenez, G.~J.~Olmo and E.~Orazi,
  Class.\ Quant.\ Grav.\  {\bf 34}, no. 23, 235003 (2017)
  doi:10.1088/1361-6382/aa9151
  [arXiv:1705.03806 [GR qc]].

\bibitem{BeltranJimenez:2019acz}
  J.~Beltran Jimenez and A.~Delhom,
  arXiv:1901.08988 [GR qc].

  \bibitem{Abbott:2017oio}
  B.~P.~Abbott {\it et al.} [LIGO Scientific and Virgo Collaborations],
  Phys.\ Rev.\ Lett.\  {\bf 119}, 141101 (2017).

  \bibitem{TheLIGOScientific:2017qsa}
  B.~P.~Abbott {\it et al.} [LIGO Scientific and Virgo Collaborations],
  Phys.\ Rev.\ Lett.\  {\bf 119}, 161101 (2017).

\bibitem{Stratta:2017igm} 
  G.~Stratta [Ligo Scientific and Virgo Collaborations],
  Nuovo Cim.\ C {\bf 40}, no. 3, 121 (2017).
  doi:10.1393/ncc/i2017-17121-7
  
\bibitem{Williams:2017ibl} 
  D.~Williams, J.~A.~Clark, A.~R.~Williamson and I.~S.~Heng,
  Astrophys.\ J.\  {\bf 858}, no. 2, 79 (2018)
  doi:10.3847/1538-4357/aab847
  [arXiv:1712.02585 [astro-ph.HE]].

  
\bibitem{Howell:2018nhu} 
  E.~J.~Howell, K.~Ackley, A.~Rowlinson and D.~Coward,
  Monthly Notices of the Royal Astronomical Society, Volume 485,
  Issue 1, May 2019, Pages 1435
  doi:10.1093/mnras/stz455
  [arXiv:1811.09168 [astro-ph.HE]].
  
\bibitem{Barack:2018yly}
  L.~Barack {\it et al.},
  arXiv:1806.05195 [GR qc].
  
  \bibitem{Dai:2019pgx} 
  Z.~G.~Dai,
  Astrophys.\ J.\  {\bf 873}, no. 2, L13 (2019)
  doi:10.3847/2041-8213/ab0b45
  [arXiv:1902.07939 [astro-ph.HE]].
  
  \bibitem{Barbieri:2019sjc} 
  C.~Barbieri, O.~S.~Salafia, A.~Perego, M.~Colpi and G.~Ghirlanda,
  Astron.\ Astrophys.\  {\bf 625}, A152 (2019)
  doi:10.1051/0004-6361/201935443
  [arXiv:1903.04543 [astro-ph.HE]].
  
\bibitem{Tarrant:2019gqg} 
  J.~Tarrant, G.~Beck and S.~Colafrancesco,
  arXiv:1904.12678 [astro-ph.HE].

\bibitem{BeltranJimenez:2017doy} 
  J.~Beltran Jimenez, L.~Heisenberg, G.~J.~Olmo and D.~Rubiera-Garcia,
  Phys.\ Rept.\  {\bf 727}, 1 (2018)
  doi:10.1016/j.physrep.2017.11.001
  [arXiv:1704.03351 [GR qc]].)

\bibitem{Olmo:2011uz} 
  G.~J.~Olmo,
  Int.\ J.\ Mod.\ Phys.\ D {\bf 20}, 413 (2011)
  doi:10.1142/S0218271811018925
  [arXiv:1101.3864 [GR qc]].

  \bibitem{Born:1934gh} 
  M.~Born and L.~Infeld,
  Proc.\ Roy.\ Soc.\ Lond.\ A {\bf 144}, no. 852, 425 (1934).
  doi:10.1098/rspa.1934.0059

\end{thebibliography}
\end{document}